\documentclass{PoS}

\usepackage{amsmath}
\usepackage{amssymb}
\usepackage{graphicx}
 \usepackage{xcolor}
\usepackage[T1]{fontenc} 
\usepackage{slashed}
\usepackage{xspace}
\usepackage{multirow}

\usepackage[utf8]{inputenc}

\newcommand{\DRbar}{{\ensuremath{\overline{\mathrm{DR}}}}}
\newcommand{\MSbar}{{\ensuremath{\overline{\mathrm{MS}}}}}
\newcommand\SARAH{{\tt SARAH}\xspace}
\newcommand{\SARAHv}[1]{\SARAH~\texttt{#1}\xspace}
\newcommand\MG{{\tt MadGraph}\xspace}

\newcommand{\MGv}[1]{\MG~{\texttt{#1}}\xspace}
\newcommand\FeynArts{{\tt FeynArts}\xspace}

\newcommand\FlavorKit{{\tt FlavorKit}\xspace}
\newcommand\FormCalc{{\tt FormCalc}\xspace}
\newcommand\CalcHep{{\tt CalcHep}\xspace}

\newcommand\WHIZARD{{\tt WHIZARD}\xspace}
\newcommand\OMEGA{{\tt O'Mega}\xspace}

\newcommand\SPheno{{\tt SPheno}\xspace}

\newcommand\FlexibleSUSY{{\tt FlexibleSUSY}\xspace}
\newcommand\Vevacious{{\tt Vevacious}\xspace}
\newcommand\MO{{\tt MicrOmegas}\xspace}
\newcommand\HB{{\tt HiggsBounds}\xspace}
\newcommand\HS{{\tt HiggsSignals}\xspace}

\newcommand\Susyno{{\tt Susyno}\xspace}
\newcommand\Mathematica{{\tt Mathematica}\xspace}

\title{Introduction to SARAH and related tools}

\ShortTitle{Introduction to SARAH and related tools}

\author{\speaker{Florian Staub}\\
        Theoretical Physics Department, CERN, Geneva, Switzerland\\
        E-mail: \email{florian.staub@cern.ch}}

\abstract{I give in this lecture an overview of the features of the \Mathematica package \SARAH, and explain how it can be used 
together with other codes to study all aspects of a BSM model. The focus will be on the description of the analytical
calculations which \SARAH can perform and how this information is used to generate automatically a spectrum 
generator based on \SPheno. I also summarize the main aspects of the other interfaces to public codes like 
\HB/\HS, \FeynArts/\FormCalc,  \CalcHep, \MO, \WHIZARD, \Vevacious or \MG.  }

\FullConference{Proceedings of the Corfu Summer Institute 2015 "School and Workshops on Elementary Particle Physics and Gravity"\\
		1-27 September 2015\\
		Corfu, Greece}

\begin{document}

\section{Introduction}
Supersymmetry (SUSY) has been the top candidate for beyond standard model (BSM) physics since many years. This has several reasons: SUSY solves the hierarchy problem of the standard model (SM), provides a dark matter candidate, leads to gauge coupling unification, and gives an explanation for electroweak symmetry breaking (EWSB). Before the LHC has turned on, the main focus has been on the minimal supersymmetric extensions of the SM, the MSSM. 
However, the negative results from SUSY searches at the LHC as well as the measured Higgs mass of about 125~GeV put large pressure on the simplest scenarios. Wide regions of the parameter space, which had been considered as natural before LHC has started, have been ruled out. This has caused more interest in non-minimal SUSY models. Also non-SUSY extensions of the SM became more popular again. Other BSM models can provide many advantages compared the MSSM:
\begin{itemize}
 \item {\bf Naturalness}: about one third of the Higgs mass has to be generated radiatively in the MSSM to explain the observation. Heavy SUSY masses are needed to gain sufficiently large loop corrections, i.e. a soft version of the hierarchy problem appears again. The need for large loop corrections gets significantly
 softened if $F$- or $D$-terms are present which already give a push to the tree-level mass.
 \item {\bf SUSY searches}: the negative results from all SUSY searches at the LHC have put impressive limits on the Sparticle masses. 
 However, the different searches are based on certain assumptions. As soon as these conditions are no longer given like in models with broken $R$-parity, the limits become much weaker. 
 \item {\bf Neutrino data}:  neutrino masses are not incorporated in the MSSM. To do that, either one of the different seesaw mechanisms can be utilised or $R$-parity 
 must be broken to allow a neutrino--neutralino mixing.
 \item {\bf Strong CP-problem}: the strong CP problems remains not only an open question in the SM but also in the MSSM. In principle, for both models the same solution exists to explain the smallness of the $\Theta$ term in QCD: the presence of a broken Peccei-Quinn symmetry.
 \item {\bf $\mu$-problem}: the superpotential of the MSSM involves one parameter with dimension mass: the $\mu$-term. This term is not protected by any symmetry, i.e. the natural values would be either exactly 0 or $O(M_{GUT})$.  However, for phenomenological reasons it should be similar to the electroweak scale.  This could be obtained if the $\mu$-term is actually not a fundamental parameter but generated dynamically like in singlet extensions. .
 \item {\bf UV completion}: it is not clear that only the gauge sector and particle content of the MSSM is the low energy limit of a GUT or String theory. Realistic UV completions predict often many additional matter at the TeV scale. In many cases also additional neutral and even charged gauge bosons are present.
 \item {\bf $R$-symmetry}: if one considers $R$-symmetric models, Majorana  masses for gauginos are forbidden. To give masses to the gauginos in these models, a coupling to a chiral superfield in the adjoint representation is needed. This gives rise to Dirac masses for the gauginos which are in agreement with $R$-symmetry. 
 Dirac gauginos are also attractive because they can weaken LHC search bounds, and flavour constraints. 
\end{itemize}

Despite the large variety and flexibility of SUSY, many dedicated public computer tools like {\tt SoftSUSY}, {\tt SPheno}, {\tt Suspect}, {\tt Isajet}  or {\tt FeynHiggs} are restricted to the simplest realization of SUSY, the MSSM, or small extensions of it. Therefore, more generic tools are needed to provide the possibility to study many other models with the same precision as the MSSM. This precision is needed to confront also these models with the strong limits from collider searches, flavour observables, dark matter observations, and Higgs measurements.
Such a tool is the \Mathematica package \SARAH. I'll given in the next section an introduction to \SARAH, before I explain the interface to \SPheno  in sec.~\ref{sec:SPheno}, to Monte-Carlo (MC) tools in sec.~\ref{sec:MC}, and to other tools in sec.~\ref{sec:tools}. I summarize in sec.~\ref{sec:summary}. Interested readers find many more details about the presented framework in Ref.~\cite{Staub:2015kfa}.

\section{BSM models and \SARAH}
\label{sec:sarah}
\subsection{Model definition}
\subsubsection{Input needed by \SARAH to define a model}
\begin{figure}[hbt]
\begin{center}
 \includegraphics[width=0.9\linewidth]{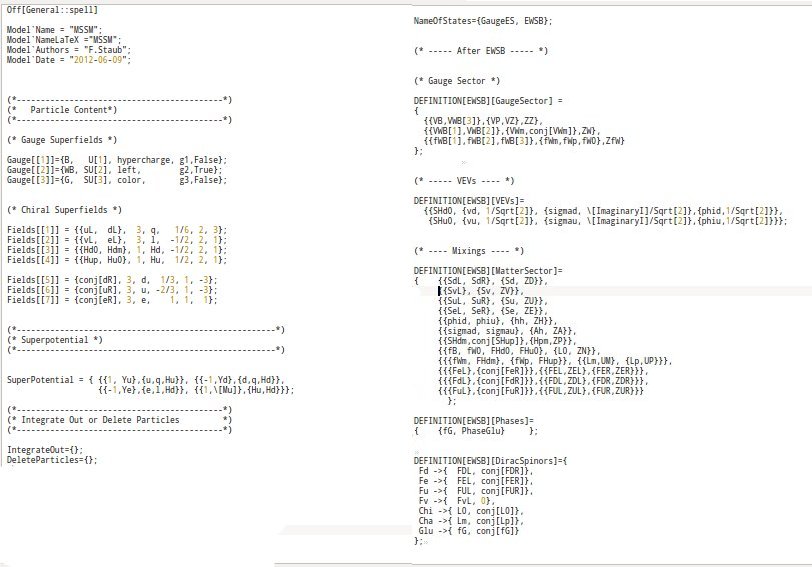}
\end{center}
\caption{The model file to implement the MSSM in \SARAH}
\label{fig:MSSM}
\end{figure}

\SARAH \cite{Staub:2008uz,Staub:2009bi,Staub:2010jh,Staub:2012pb,Staub:2013tta} is optimized for the handling of a wide range of BSM models. The basic aim of \SARAH is to give the user the possibility to implement models in an easy, compact and straightforward way. Most tasks to get the Lagrangian are fully  automatized; it is sufficient to define just the fundamental properties of the model:
\begin{enumerate}
 \item Global symmetries
 \item Gauge symmetries
 \item Chiral superfields
 \item (Super)potential
 \item Field rotations
\end{enumerate}
That means that \SARAH automatizes many steps to derive the Lagrangian from that input:
\begin{enumerate}
 \item All interactions of matter fermions and the $F$-terms are derived from the (super)potential
 \item All vector boson and gaugino interactions as well as $D$-terms are derived from gauge invariance
 \item All gauge fixing terms are derived by demanding that scalar--vector mixing vanishes in the kinetic terms
 \item All ghost interactions are derived from the gauge fixing terms
 \item All soft-breaking masses for scalars and gauginos as well as the soft-breaking counterparts to the superpotential couplings are added automatically  for SUSY models
 \item All rotations from the gauge to the mass eigenstates are performed
\end{enumerate} 
The range of models which can be handled by \SARAH is very broad: 
\begin{itemize}
\item {\bf Global symmetries}: \SARAH can handle an arbitrary number of global symmetries which are either $Z_N$ or $U(1)$ symmetries. Also a continuous $R$-symmetry $U(1)_R$ is possible.  
\item {\bf Gauge sector}: \SARAH is not restricted to the SM gauge sector, but many more gauge groups can be defined. To improve the power in dealing with gauge groups, \SARAH has linked routines from the \Mathematica package \Susyno \cite{Fonseca:2011sy}. \SARAH together with \Susyno take care of all group-theoretical calculations: the Dynkin and Casimir invariants are calculated, and the needed representation matrices as well as Clebsch-Gordan coefficients are derived. Also gauge kinetic mixing is fully supported for an arbitrary number of Abelian gauge groups. 
\item {\bf Matter sector}: there is no practical restriction how many matter states --either defined as chiral superfields in SUSY models, are as component fields in non-SUSY models-- are present. All states can come with an arbitrary number of generations and can transform as any irreducible representation with respect to the defined gauge groups.
\item {\bf (Super)potential}: the main restriction is that only renormalizable terms are supported in the (super)potential. However, there is no restriction how many terms are present.
\end{itemize}
As example how short a model implementation in \SARAH is, I show in Fig.~\ref{fig:MSSM} the model file for the MSSM. 

\subsubsection{Checks of implemented models}
\label{sec:checks}
\SARAH provides functions to check the (self-) consistency of a model addressing the following questions:
\begin{itemize}
 \item Causes the particle content gauge anomalies?
 \item Leads the particle content to the Witten anomalies?
 \item Are all terms in the (super)potential in agreement with global and local symmetries?
 \item Are there other terms allowed in the (super)potential by global and local symmetries?
 \item Are all unbroken gauge groups respected?
 \item Are there terms in the Lagrangian of the mass eigenstates which can cause additional mixing between fields?
 \item Are all mass matrices irreducible?
 \item Are the properties of all particles and parameters defined correctly?
\end{itemize}

\subsection{Calculations and output}
\label{sec:analytical}
\SARAH can perform in its natural \Mathematica environment many calculations for a model on the analytical level. For an exhaustive numerical analysis usually one of the dedicated interfaces  to other tools discussed in the next sections is the best approach. I start with some details about the analytical calculations which \SARAH can perform. 

\subsubsection{Tree-level properties}
\paragraph{Tadpole equations} During the evaluation of a model, \SARAH calculates 'on the fly' all minimum conditions of the tree-level potential, the so called tadpole equations.

\paragraph{Masses} \SARAH calculates for all states which are rotated to mass eigenstates the mass matrices during the evaluation of a model. In addition, it calculates also the masses for the states for which no field rotation has taken place. 

\paragraph{Vertices} \SARAH has functions to extract in an efficient way all tree-level vertices from the Lagrangian. These vertices are saved in different \Mathematica arrays according to their generic type. 

\subsubsection{Renormalization group equations}
\label{sec:RGEs}
\SARAH calculates for SUSY and non-SUSY models the full two-loop RGEs including the full CP and flavour structure. I give here a short summary of the underlying, generic results which are used in these calculations.  
\paragraph*{SUSY RGEs} The calculation of the SUSY RGEs is mainly based on Ref.~\cite{Martin:1993zk}. However, this work does not cover all possible subtleties which can appear in SUSY models. Therefore, \SARAH has implemented also some more results from literature which became available in the last few years: 
\begin{itemize}
 \item In the case of several $U(1)$'s, gauge-kinetic mixing can arise if the groups are not orthogonal. Substitution rules to translate the results of Ref.~\cite{Martin:1993zk} to those including gauge kinetic mixing where presented in Ref.~\cite{Fonseca:2011vn} and are used by \SARAH 
 \item The calculation of the RGEs in the presence of Dirac gaugino is based on Ref.~\cite{Goodsell:2012fm}
 \item The results of Refs.~\cite{Sperling:2013eva,Sperling:2013xqa} are used to get the gauge dependence in the running of the VEVs.
\end{itemize}

\paragraph*{Non-SUSY RGEs} \SARAH uses the expressions of Refs.~\cite{Machacek:1983tz,Machacek:1983fi,Machacek:1984zw,Luo:2002ti} for the calculation of the RGEs in a general quantum field theory. These results are completed by Ref.~\cite{Fonseca:2013bua} to cover gauge kinetic mixing and again by Refs.~\cite{Sperling:2013eva,Sperling:2013xqa} to include the gauge-dependence of the running VEVs also in the non-SUSY case.

\subsubsection{One- and two-loop corrections to tadpoles and self-energies}
\label{sec:loopcorrections}
\begin{figure}[hbt]
\centering
\includegraphics[width=1.0\linewidth]{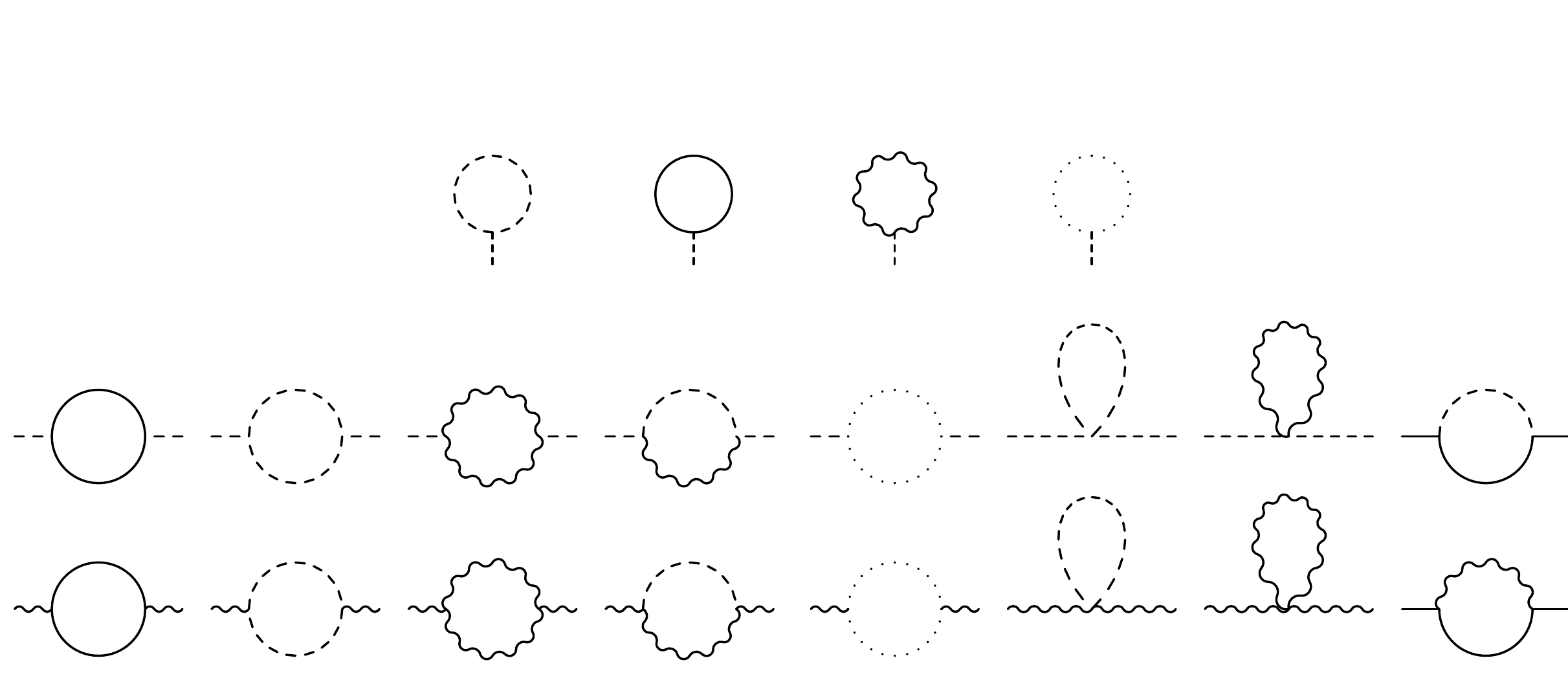} 
\caption{Generic diagrams included by \SARAH to calculate one-loop tadpoles and self-energies.}
\label{fig:1loopDiagrams}
\end{figure}

\paragraph{One-loop corrections} \SARAH calculates the analytical expressions for the one-loop corrections to the tadpoles and the one-loop self-energies for all particles. For states which are a mixture of several gauge eigenstates, the self-energy matrices are calculated. The calculations are performed in $\DRbar$-scheme using 't Hooft gauge. In the case of non-SUSY models \SARAH switches to $\MSbar$-scheme. All generic diagrams which are covered at one-loop are depicted in Fig.~\ref{fig:1loopDiagrams}. 

\paragraph{Two-loop corrections} It is even possible to go beyond one-loop with \SARAH and to calculate two-loop contributions to the self-energies of real scalars. There are two equivalent approaches implemented in the \SPheno interface of \SARAH to perform these calculations: an effective potential approach, and a diagrammatic approach with vanishing external momenta. More details about these calculations are given in sec.~\ref{sec:TwoLoop}.

\subsection{Output}
The information derived by \SARAH can be outputted in different ways
\begin{itemize}
 \item Model files for Monte-Carlo tools like \CalcHep, \MG, or \WHIZARD
 \item Model files for other tools like \FeynArts and \Vevacious 
 \item Fortran code to implement a model in \SPheno \footnote{A similar interface to generate a spectrum generator written in {\tt C+} became available via \FlexibleSUSY \cite{Athron:2014yba}.}. 
 \item \LaTeX\ files 
\end{itemize}
A few more details about the different outputs are given in the next three sections. 

\section{Linking \SPheno and \SARAH}
\label{sec:SPheno}
\SARAHv{3} has been the first `spectrum--generator--generator': SARAH writes Fortran source code
for \SPheno using the derived information about the 
mass matrices, tadpole equations, vertices, loop corrections and RGEs for the given model. 
Based on this code the user gets a fully functional spectrum generator for a model. 
The features of a spectrum generator created in this way are 
\begin{itemize}
\item Full {two-loop running} of all parameters and {all masses at one-loop}
\item {two-loop} corrections to {Higgs} masses
\item Complete {one-loop thresholds} at $M_Z$
\item calculation of {flavour} and {precision observables} at full 1-loop
\item calculation of {decay widths} and {branching ratios}
\item interface to {\tt HiggsBounds} and {\tt HiggsSignals}
\item estimate of electroweak {Fine-Tuning}
\end{itemize}
I give a few more details about the different calculations. 

\subsection{Mass calculation with \SPheno}
\subsubsection{Threshold corrections}
For a precise calculation of the masses, it is necessary to have an accurate input for all parameters which enter the calculation. In general, the running SM parameters depend on the masses of BSM states. The reason are the thresholds to match the running parameters to the measured ones. The routines generated by \SARAH perform a full one-loop matching in the given model to calculate the SM gauge and Yukawa couplings. 

\subsubsection{One-loop masses}
The one-loop mass spectrum is calculated from the information about the one-loop self-energies and tadpole equations. The procedure is a generalisation of the one explained in detail for the MSSM in Ref.~\cite{Pierce:1996zz}. The main features are 
\begin{itemize}
 \item Any one-loop contribution in a given model to all fermions, scalars and vector bosons is included 
 \item The full $p^2$ dependence in the loop integrals is included
 \item An iterative procedure is applied to find the on-shell masses $m(p^2=m^2)$. 
\end{itemize}

\subsubsection{Two-loop Higgs masses}
\label{sec:TwoLoop}
\begin{figure}[hbt]
 \includegraphics[width=1.0\linewidth]{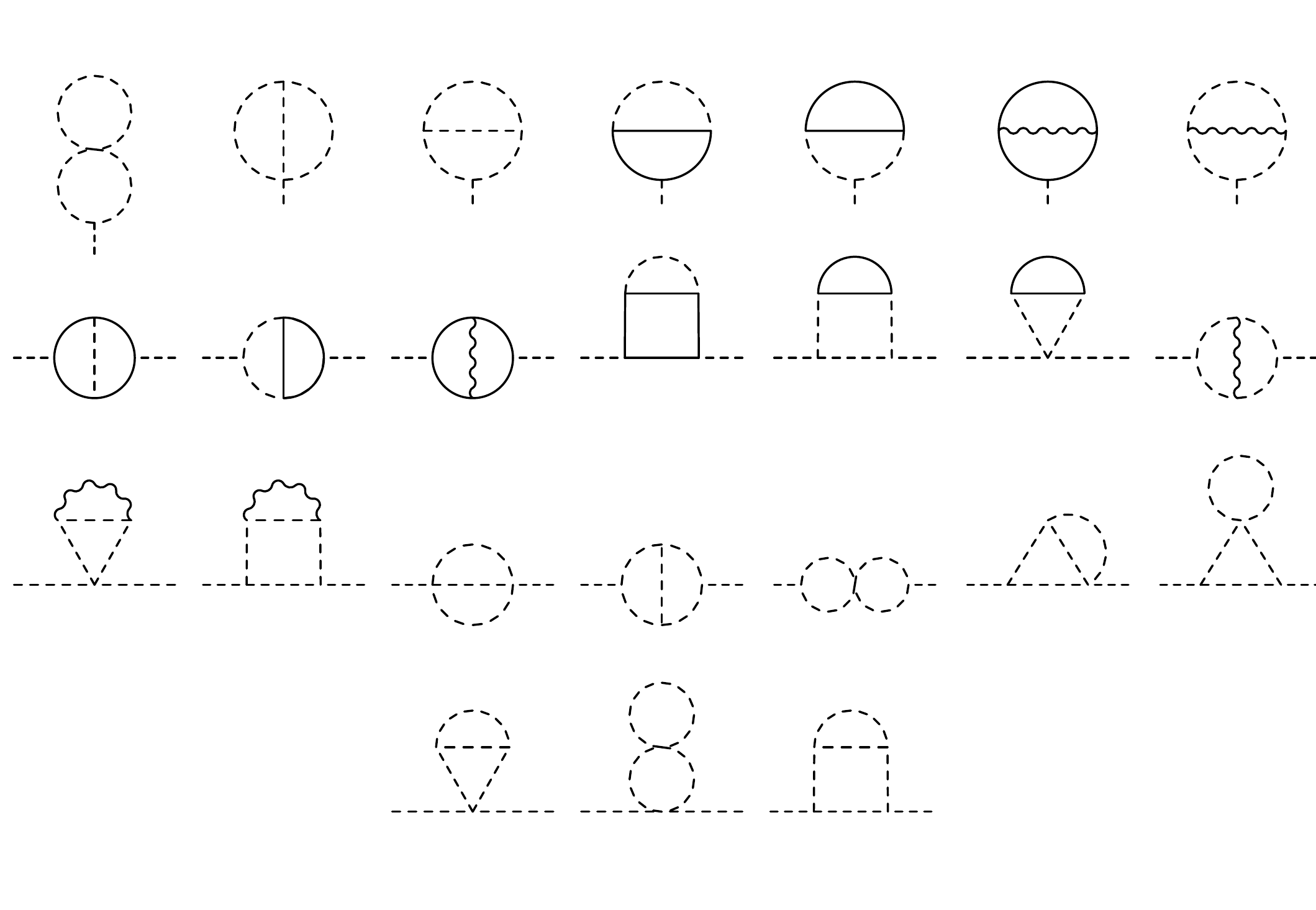}
 \caption{Generic diagrams included to calculate the two two-loop corrections to tadpoles and scalars. These are the diagram which don't vanish in the gaugeless.}
\label{fig:2loopDiagram}
\end{figure}

\begin{figure}[hbt]
\includegraphics[width=0.45\linewidth]{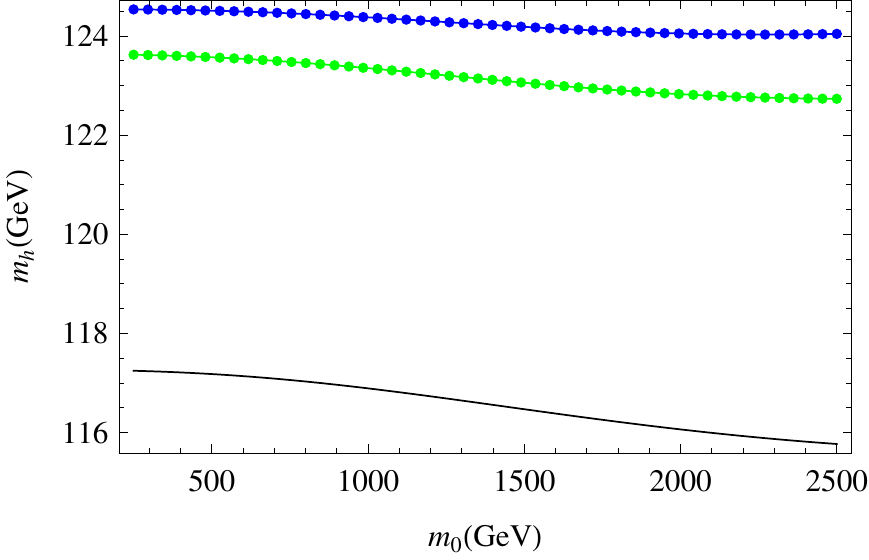} \hfill 
  \includegraphics[width=0.45\linewidth]{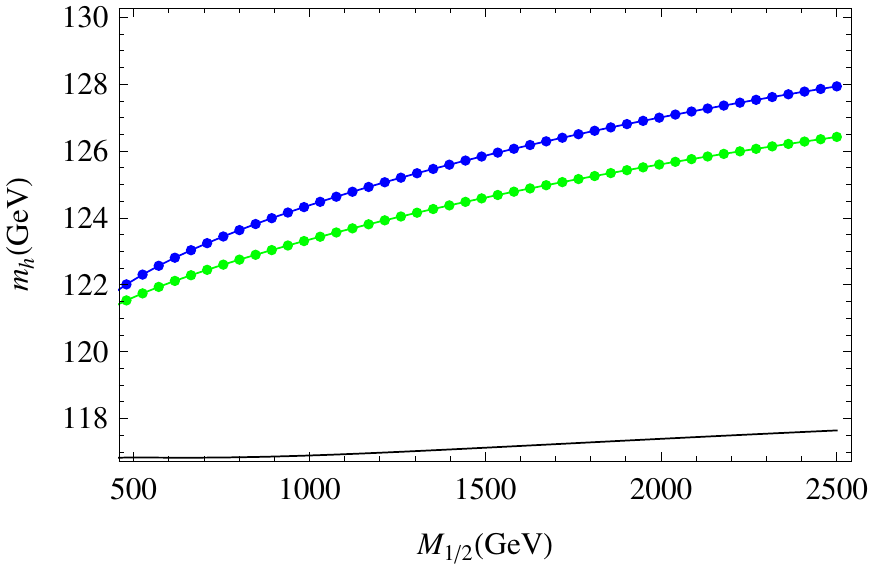} \\ 
 \includegraphics[width=0.45\linewidth]{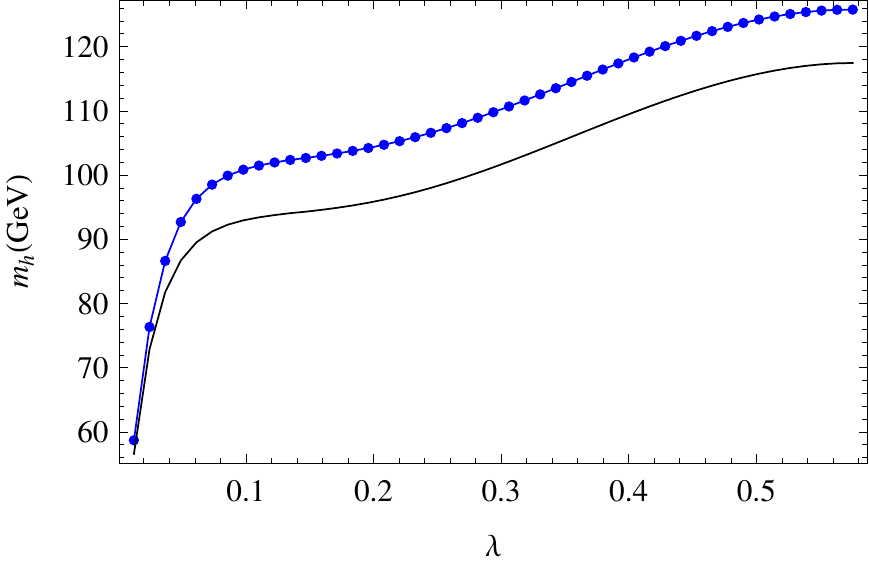} \hfill 
 \includegraphics[width=0.45\linewidth]{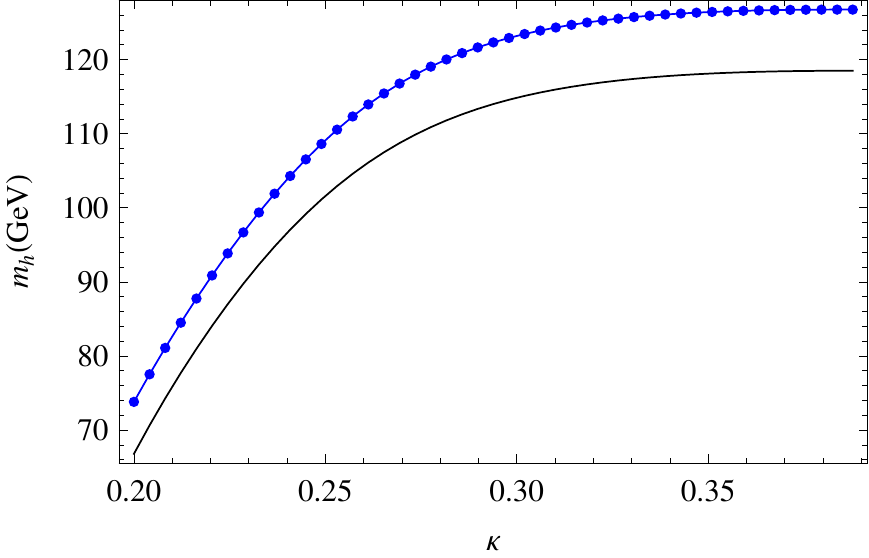}
 \caption{Comparison of the Higgs mass calculation at two-loop level using the automatically generated routines of \SARAH (full lines) and the routines based in 
 Refs.~\cite{Brignole:2001jy,Degrassi:2001yf,Brignole:2002bz,Dedes:2002dy,Dedes:2003km} for the MSSM and Ref.~\cite{Degrassi:2009yq} for the NMSSM. The green lines include only the corrections $O(\alpha_s (\alpha_t+\alpha_b))$, while for the blue lines all other, available corrections are included. The black line corresponds to the one-loop results.}
\label{fig:TwoLoop} 
\end{figure}

\SARAH can also generate Fortran code to calculate the two-loop corrections to real scalars with \SPheno. The same approximation as usual done for the MSSM are also applied here: (i) all calculations are performed in the gaugeless limit, i.e. the electroweak contributions are dropped, and  (ii) the momentum dependence is neglected. Using these routines,the theoretical uncertainty in the Higgs mass prediction for many models has shrunken now to the level of the MSSM. In general, there are two different techniques to calculate the two-loop corrections with \SARAH\SPheno:
\begin{itemize}
 \item {\bf Effective potential calculation}: \SARAH makes use of the generic two-loop results for the effective potential given in Ref.~\cite{Martin:2001vx}. To get the values for the two-loop self-energies and two-loop tadpoles, the derivatives of the potential with respect to the VEVs are taken numerically as proposed in Ref.~\cite{Martin:2002wn}. There are two possibilities for this derivation implemented in \SARAH/\SPheno: (i)  a fully numerical procedure which takes the derivative of the full effective potential with respect to the VEVs. A semi-analytical derivation which takes analytical the derivative of the loop functions with respect to involved masses, but derives the masses and coupling numerically with respect to the VEVs. More details about both methods and the numerical differences are given in Ref.~\cite{Goodsell:2014bna}. 
 \item {\bf Diagrammatic calculation}: A fully diagrammatic calculation for two-loop contributions to scalar self-energies with \SARAH--\SPheno became available with Ref.~\cite{Goodsell:2015ira}. In this setup a set of generic expressions first derived in  Ref.~\cite{Goodsell:2015ira} is used. All two-loop diagrams shown in Fig.~\ref{fig:2loopDiagram} are included in the limit $p^2=0$. The advantage of the diagrammatic approach is that no numerical derivation is needed.
\end{itemize}
A comparison between the results using the new routines generated by \SARAH and widely used routines based on Refs.~\cite{Brignole:2001jy,Degrassi:2001yf,Brignole:2002bz,Dedes:2002dy,Dedes:2003km} for the MSSM and Ref.~\cite{Degrassi:2009yq} for the NMSSM is shown in Fig.~\ref{fig:TwoLoop}. Obviously, a perfect agreement is found for both models.   
Since there are no further possibilities to cross check the results obtained for the Higgs mass with \SARAH/\SPheno, the above described implementation of two independent approaches (effective potential, diagrammatic) was necessary to have the possibility to double check results. In the meantime, this setup has been used to analyse the Higgs mass also in other non-minimal SUSY models at the two-loop level:
\begin{itemize}
 \item Contributions of trilinear RpV, see Ref.~\cite{Dreiner:2014lqa}
 \item Missing corrections for NMSSM, see Ref.~\cite{Goodsell:2014pla}
 \item Contributions from non-holomorphic soft-terms, see Ref.~\cite{Un:2014afa}
 \item MRSSM at full two-loop, see Ref.~\cite{Diessner:2015yna}
 \item Contributions from vectorlike stops, see Ref.~\cite{Nickel:2015dna}
\end{itemize}

\subsection{Decay widths and branching ratios}
\SPheno modules created by \SARAH calculate all two-body decays for SUSY and Higgs states as well as for additional gauge bosons. In addition, the three-body decays of a fermion into three other fermions, and of a scalar into another scalar and two fermions are included. \\
In the Higgs sector, possible decays into two SUSY particles, leptons and massive gauge bosons are calculated at tree-level. For two quarks in the final state the dominant QCD corrections due to gluons are included \cite{Spira:1995rr}. The loop induced decays into two photons and gluons are fully calculated at leading-order (LO) with the dominant next-to-leading-order corrections known from the MSSM. For the LO contributions all charged and coloured states in the given model are included in the loop, i.e. new contributions rising in a model beyond the MSSM are fully covered at one-loop. In addition, in the Higgs decays also final states with off-shell gauge bosons ($Z Z^*$, $W W^*$) are included. 

\subsection{Flavour observables}
\begin{figure}[hbt]
 \includegraphics[width=0.45\linewidth]{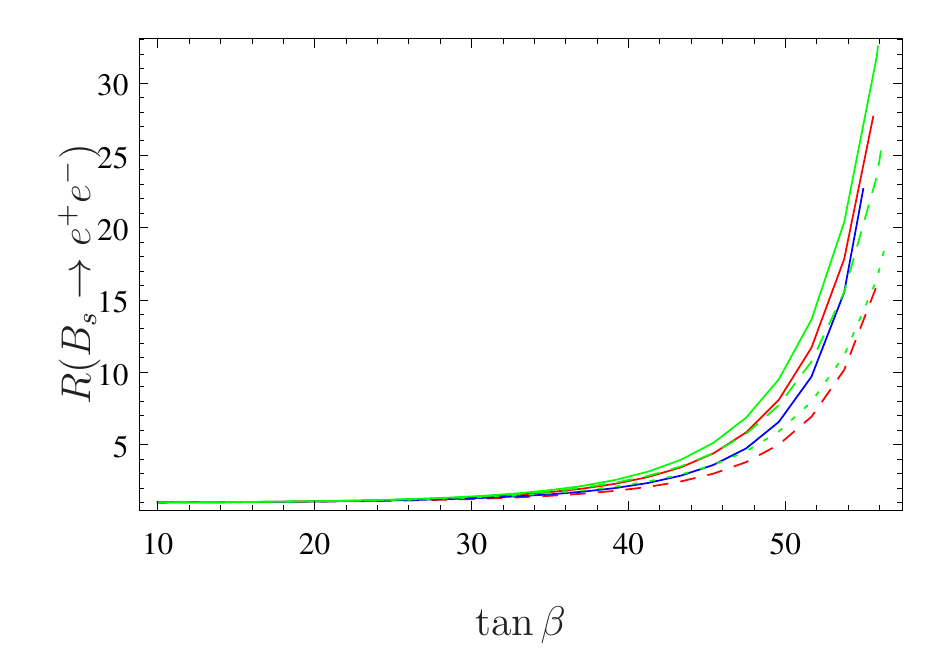}
 \hfill 
 \includegraphics[width=0.45\linewidth]{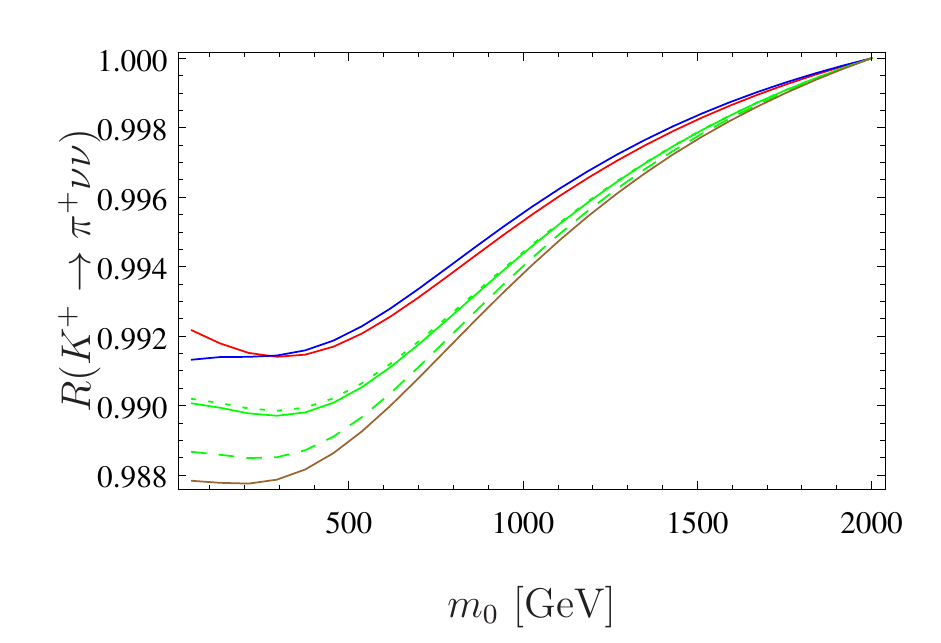} \\
 \includegraphics[width=0.45\linewidth]{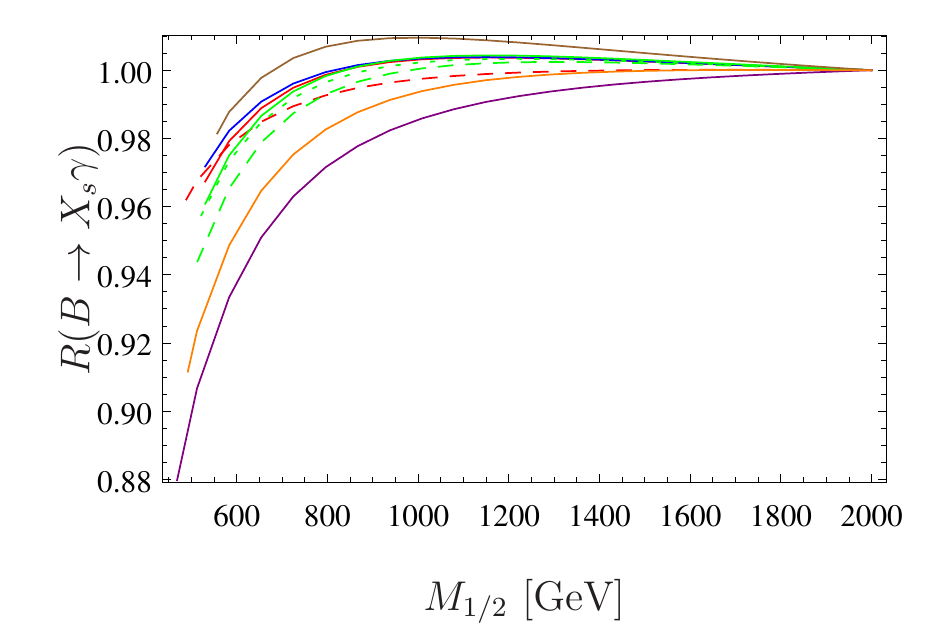}
 \hfill 
 \includegraphics[width=0.45\linewidth]{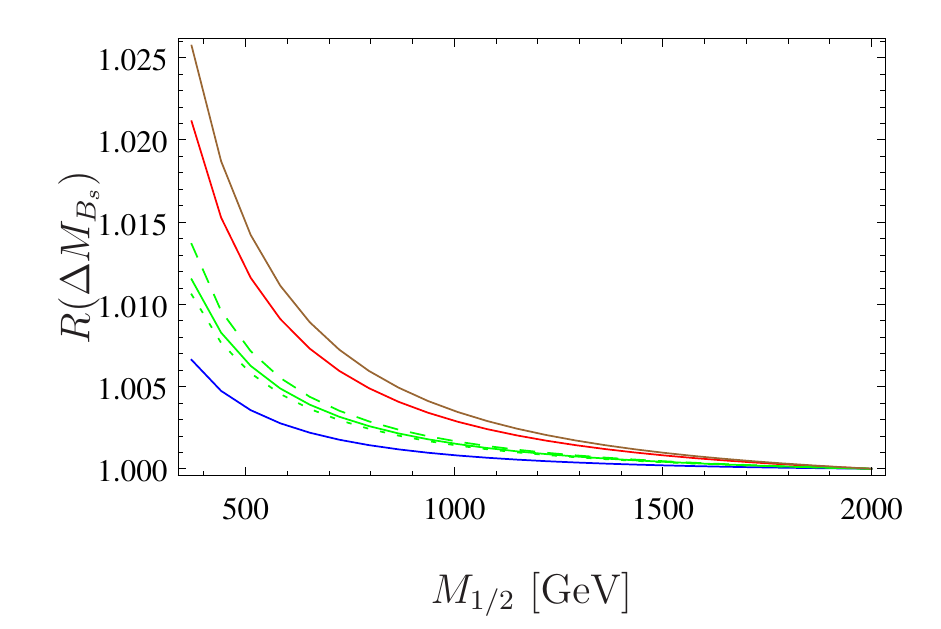}  
\caption{Comparison of different flavour observables in the CMSSM. The colour code is: {\tt FlavorKit} (red), {\tt SPheno 3.3} (blue), {\tt SUSY\_Flavor 1} (brown), {\tt SUSY\_Flavor 2} (green), {\tt MicrOmegas} (orange), {\tt SuperIso} (plum).}
\label{fig:flavour} 
\end{figure}

The \SPheno modules written by \SARAH contain already out of the box the routines to calculate many quark and lepton flavour violating observables:
\begin{itemize}
 \item Lepton flavour violation:
\begin{itemize}
  \item Br($l_i \to l_j \gamma$), Br($l \to 3l'$), Br($Z\to l l'$)
  \item CR($\mu-e,N$) { (N=Al,Ti,Sr,Sb,Au,Pb)}, Br($\tau\to l+P$) {  (P=$\pi$, $\eta$,$\eta'$)}
 \end{itemize} 
 \item Quark flavour violation:
\begin{itemize}
  \item Br($B\to X_s\gamma$), Br($B_{s,d}^0 \to l \bar{l}$), Br($B \to s l \bar{l}$), Br($K \to \mu \nu$)
  \item Br($B \to q \nu\nu$), Br($K^+ \to \pi^+ \nu\nu$), Br($K_L \to \pi^0 \nu\nu$)
  \item $\Delta M_{B_s,B_d}$, $\Delta M_K$, $\epsilon_K$, Br($B \to K \mu \bar{\mu}$)
  \item Br($B\to l \nu$), Br($D_s \to l \nu$)
 \end{itemize} 
\end{itemize}
The calculation is based on the \FlavorKit functionality \cite{Porod:2014xia} which is a connection  \FeynArts--\SARAH--\SPheno. This provides a full one-loop calculation in a given model. In addition, this interface can be used to derive Wilson coefficients for new operators, and to relate coefficients to calculate new observables with \SPheno. In order to demonstrate that this fully automatized calculation reproduces the results from dedicated MSSM codes, a comparison is shown in Fig.~\ref{fig:flavour}.

\subsection{Checking Higgs constraints}
\HB \cite{Bechtle:2008jh,Bechtle:2011sb,Bechtle:2013wla} and \HS \cite{Bechtle:2013xfa} are dedicated tools to study the Higgs properties of a given parameter point in a particular model. While \HB checks the parameter point against exclusion limits from Higgs searches, \HS gives a $\chi^2$ value to express how good the point reproduces the Higgs measurements. In general, \HS and \HB can handle different inputs: either the cross sections for all necessary processes can be given at the parton or hadron level, or the effective couplings of the Higgs states to all SM particles are taken as input. In addition, the masses and widths of all neutral as well as charged Higgs states are always needed. 
\SPheno provides all input for the effective coupling approach. The information is given in the SLHA spectrum file and in addition in separated files (called {\tt MH\_GammaTot.dat}, {\tt  MHplus\_GammaTot.dat}, {\tt BR\_H\_NP.dat}, {\tt BR\_Hplus.dat}, {\tt BR\_t.dat}, {\tt effC.dat}). While SLHA files can be used with \HB for models with up to five neutral scalars, the  separated files can be used with even up to nine neutral and nine charged scalars.

\section{MC Tools and \SARAH}
\label{sec:MC}
\SARAH writes all necessary files to implement a new model in different MC tools. I give a short overview of the main features and explain, how parameter values between \SPheno and the MC can be exchanged. 
\subsection{Output of models files}
\paragraph{\CalcHep}
The model files for \CalcHep \cite{Pukhov:2004ca,Boos:1994xb} can also be used with \MO \cite{Belanger:2014hqa} for dark matter studies. 
\CalcHep is  able to read the numerical
 values of the masses and mixing matrices from a SLHA spectrum file based on the 
 \texttt{SLHA+} functionality \cite{Belanger:2010st}. \SARAH makes use of this functionality to generate the model files in a way that they read the spectrum files written by a \SARAH generated \SPheno version.

\paragraph{UFO format}
\SARAH can write model files in the UFO format \cite{Degrande:2011ua} to implement new models in
 \MGv{5}. This format is also supported by other tools like {\tt GoSam}
 \cite{Cullen:2011ac}, {\tt Aloha} \cite{deAquino:2011ub}, 
 \texttt{Herwig++} \cite{Gieseke:2003hm,Gieseke:2006ga} or \texttt{MadAnalysis 5} \cite{Conte:2012fm}. \\
One can give the spectrum file written by \SPheno as parameter card to \MG.

\paragraph{\WHIZARD/\OMEGA}
\SARAH writes all necessary files to implement a model in \WHIZARD and \OMEGA  \cite{Kilian:2007gr,Moretti:2001zz}. Since the SLHA reader of \WHIZARD is at the moment restricted to the MSSM and NMSSM, \SPheno versions generated by \SARAH write all information about the spectrum and parameters also in an additional file in the \WHIZARD specific format. This file can be then read by \WHIZARD. 

\subsection{Interplay \SARAH--\SPheno--MC-Tool} 
The tool chain \SARAH--\SPheno--MC-Tools has one very appealing feature: the implementation of a model in the spectrum generator (\SPheno) as well as in a MC tool is based on just one single implementation of the model in \SARAH. Thus, the user has not to worry that the codes might use different conventions to define the model. In addition, \SPheno also provides all widths for the particles and this information can be used by the MC-Tool as well to save time.

\section{Other output}
\label{sec:tools}

\subsection{\FeynArts/\FormCalc}
\FeynArts/\FormCalc   \cite{Hahn:2000kx,Hahn:2009bf} is a very powerful combination of two \Mathematica packages to perform tree-level and loop calculations. \FeynArts is a Mathematica package for the {generation and visualization} of Feynman {diagrams} and {amplitudes}. \FormCalc makes use of Feynman diagrams generated by \FeynArts, and of {\tt Form} for {analytical manipulations}: it calculates the amplitude, squares it, performs {colour simplifications}, makes the {polarisation sums}, and so on. \FormCalc can also write {\tt Fortran} or {\tt C} code for further {numerical calculations}. \SARAH can be used to generate the model files for these tools to implement a new model.

\subsection{\Vevacious}
\Vevacious  is a tool to check for the global minimum of the one-loop effective potential for a given model allowing for a particular set of non-zero VEVs. For this purpose \Vevacious finds first all tree-level minima by using {\tt HOM4PS2} \cite{lee2008hom4ps}. Afterwards, it minimizes the one-loop effective potential starting from these minima using {\tt minuit}  \cite{James:1975dr}. If the input minimum turns out not to be the global one, life-time of meta-stable vacua can be calculated using {\tt Cosmotransitions} \cite{Wainwright:2011kj}.\\
\Vevacious takes the tadpole equations, the polynomial part of the  scalar potential and all mass 
matrices as input. All of this information has to be expressed including all VEVs which should be tested. That means, in order to check for charge and colour breaking minima the stop and stau have to show up in the potential and mass matrices and the entire mixing triggered by these VEVs should be included. To take care of all that, the corresponding input files can be generated by \SARAH. Possible applications when these checks become important are:
\begin{itemize}
 \item Charge and colour breaking minima, see for instance Refs.~\cite{Camargo-Molina:2013sta,Camargo-Molina:2014pwa}
 \item Spontaneous R-parity violation, see for instance Ref.~\cite{CamargoMolina:2012hv} 
 \item NMSSM with large $A$ terms, see for instance Refs.~\cite{Kanehata:2011ei,Kobayashi:2012xv}
\end{itemize}

\subsection{\LaTeX}
All derived information (mass matrices, vertices, RGEs, loop corrections) can be exported to 
\LaTeX\ to get the expressions in a human readable form. 
There exists the possibility to draw Feynman diagrams with {\tt FeynMF} \cite{Ohl:FeynMF} 
for all vertices. 

\section{Summary}
\label{sec:summary}
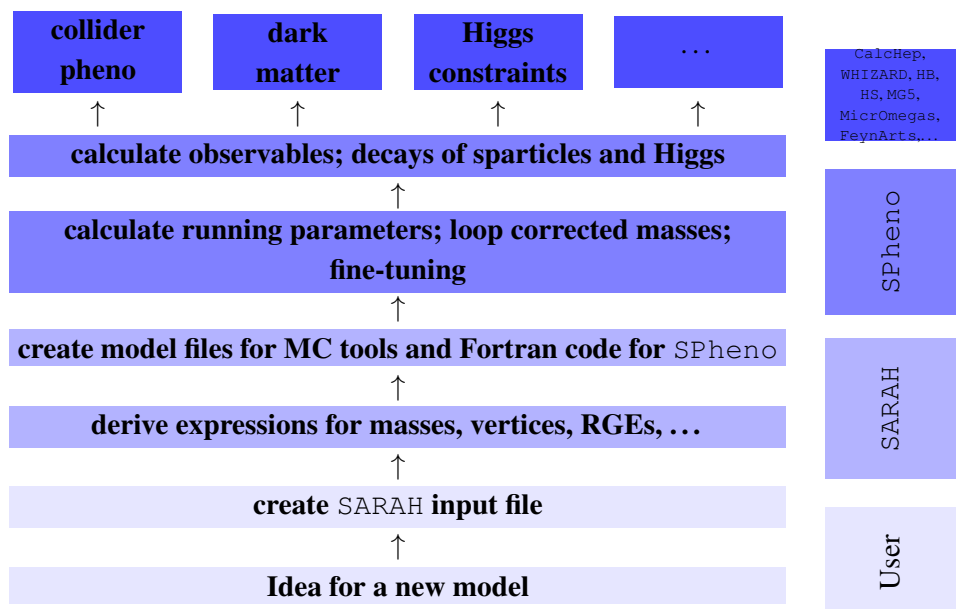
\begin{figure}[hbt]
\begin{minipage}{\linewidth}
\begin{center}
\begin{tabular}{cccc}
\colorbox{blue!70}{\parbox{2.cm}{\centering \bf collider pheno}} &
\colorbox{blue!70}{\parbox{2.cm}{\centering \bf dark\\ matter}} &
\colorbox{blue!70}{\parbox{2.cm}{\centering \bf Higgs constraints}}  &
\colorbox{blue!70}{\parbox{2.cm}{\centering \vspace{0.35cm} \dots
\\\vspace{0.35cm}}}   \\ 
\(\uparrow\) & \(\uparrow\) &  \(\uparrow\) &  \(\uparrow\) \\
\end{tabular}
\colorbox{blue!50}{\parbox{10cm}{\centering \bf calculate observables; decays of sparticles and Higgs}} \\
\(\uparrow\)\\
\colorbox{blue!50}{\parbox{10cm}{\centering {\bf calculate running parameters; loop corrected masses; fine-tuning}}} \\
\(\uparrow\)\\
\colorbox{blue!30}{\parbox{10cm}{\centering {\bf create model files for MC tools and Fortran code for \SPheno}}} \\
\(\uparrow\)\\
\colorbox{blue!30}{\parbox{10cm}{\centering {\bf derive expressions for masses, vertices, RGEs, \dots}}} \\
\(\uparrow\)\\
\colorbox{blue!10}{\parbox{10cm}{\centering {\bf create \SARAH input file}}} \\
\(\uparrow\)\\
\colorbox{blue!10}{\parbox{10cm}{\centering {\bf Idea for a new model}}}
\end{center} 
\begin{picture}(0,0)
\put(375,18){\mbox{\rotatebox{90}{\colorbox{blue!10}{\parbox[c][1.5cm]{1.24cm}{\centering User}}}}}
\put(375,70){\mbox{\rotatebox{90}{\colorbox{blue!30}{\parbox[c][1.5cm]{1.65cm}{\centering \SARAH}}}}}
\put(375,132){\mbox{\rotatebox{90}{\colorbox{blue!50}{\parbox[c][1.5cm]{1.72cm}{\centering \SPheno}}}}}
\put(375,212){\mbox{\rotatebox{0}{\colorbox{blue!70}{\parbox[c][1.00cm]{1.5cm}{\centering \tiny \CalcHep, \WHIZARD, {\tt HB}, {\tt HS}, {\tt MG5},  \MO, \FeynArts,\dots}}}}}
\end{picture}
\end{minipage} 
\caption{Schematic framework to study all aspects of a new model in a highly automatized way.}
\label{fig:summary}
\end{figure}

I have given in this lecture a short summary how the study of a new particle physics models can be automatized to a large extent. The first step is to implement a new model in \SARAH. \SARAH then calculates all analytical properties of this model and passes this information to other tools for the numerical evaluation. The main work-flow of this setup is summarized in Fig.~\ref{fig:summary}. 

\section*{Acknowledgements}
I thank the organizers of ``Summer School and Workshop on the Standard Model and Beyond'', Corfu2015, for the invitation and the hospitality during the stay.

\end{document}